# RECONFIGURATION STRATEGIES FOR ONLINE HARDWARE MULTITASKING IN EMBEDDED SYSTEMS

Marcos Sanchez-Elez and Sara Roman

[1]Department of Arquitectura de Computadores, Universidad Complutense, Madrid, Spain
marcos@fis.ucm.es sroman@dacya.ucm.es

## ABSTRACT

*An intensive use of reconfigurable hardware is expected in future embedded systems. This means that the system has to decide which tasks are more suitable for hardware execution. In order to make an efficient use of the FPGA it is convenient to choose one that allows hardware multitasking, which is implemented by using partial dynamic reconfiguration. One of the challenges for hardware multitasking in embedded systems is the online management of the only reconfiguration port of present FPGA devices. This paper presents different online reconfiguration scheduling strategies which assign the reconfiguration interface resource using different criteria: workload distribution or task' deadline. The online scheduling strategies presented take efficient and fast decisions based on the information available at each moment. Experiments have been made in order to analyze the performance and convenience of these reconfiguration strategies.*

## 1. INTRODUCTION

The combined need for flexibility and high performance in embedded systems today is clearly pointing at the use of Reconfigurable Devices as the new design paradigm [1]. This expectation is based on the performance gain achieved by using a FPGA because of the possibility to dynamically reconfigure parts of the FPGA during run-time without disturbing the execution of the other parts [3].

In order to achieve the maximum processing performance, in the system that combines reconfigurable hardware and microprocessor, either the operating system (OS) or the application designer, selects the tasks that are most suitable for hardware execution, and distribute the workload between the FPGA and the microprocessor [2]. In this case the reconfiguration time is an aspect of FPGA technology which adds a significant overhead to hardware execution and becomes very important for an efficient use of present day FPGAs, as they only have one reconfiguration port.

In such hybrid embedded designs, the decision on whether a task will be executed in the microprocessor or the FPGA is taken during execution time. This means that the tasks that are going to be executed are not known in advance: the multitasking management is an online problem. Therefore, in this approach, it is not possible to use a known task graph to schedule task reconfiguration and execution in the FPGA. In addition, once a task begins executing in the FPGA it is not feasible to interrupt its execution: saving a hardware context is complex and time consuming for the implementation of an effective real-time computing system. For these reasons we are using a non-preemptive online model.

There are researchers, as is discussed in Related Work Section, have developed offline algorithms to scheduling the reconfiguration of tasks. This approach is not completely applicable to a real-time multitasking problem. On the other hand, many researchers have developed online heuristics mainly focused on optimizing FPGA area use and where the impact of the reconfiguration scheduling on the overall scheduling of tasks is neglected by implicitly including reconfiguration time in the task execution time.

Scheduling the execution of tasks on the only basis of free available space leads to erroneous task scheduling, as shown in the example in Fig. 1, in which several tasks have been scheduled for execution following a First Fit area use heuristic. Fig. 1 presents a snapshot of the execution of several tasks scheduling with an algorithm that considers the reconfiguration time implicitly included in the execution time of a task (shown as trec + tex). When a task arrives at the FPGA it is scheduled for immediate execution if there is enough free space in the FPGA. In the example, when task T3 arrives at the FPGA there is enough free space and this task is scheduled for execution from time units 4 to 17. However T2, which arrived previously, is being reconfigured from time units 3 to 6, which means that both tasks would be reconfigured at the same time. This error in the scheduling is caused by the lack of an explicit difference between execution and reconfiguration times. This situation is remarked in the figure by red line rectangles.

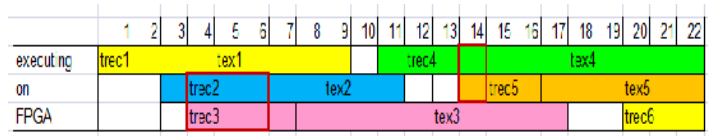

Fig 1.a Task scheduling without reconfiguration scheduling

This example shows that, in order to provide a correct task scheduling, a real-time algorithm needs to take into account the free area in the device as well as the availability of the reconfiguration port, and that the real problem is an online problem. However, in a hardware multitasking system with a high task frequency rate, this simple scheduling heuristic is not the most efficient, as proved in this paper. When incoming tasks may not be immediately sent for execution in the FPGA, one or more waiting queues may be set and other strategies to schedule the execution of pending tasks in the FPGA may be applied.

The objective of this paper is to present different real time online reconfiguration strategies and to compare them in order to select the best one. We have compared a simple strategy, like the one shown in the example, with another strategy which makes an intensive use of the reconfiguration interface.

The rest of this paper is organized as follows: section 2 presents the previous work in this subject, section 3 describes the system model used for this study, section 4 the basic reconfiguration algorithm, section 5 presents the Reconfiguration Intensive Use Strategy implemented using two different priority policies. Section 6 shows the experimental results obtained and section 7 provides conclusions about the different reconfiguration strategies which are the subject of this study.

## 2. STATE OF THE ART

We may find numerous papers related to task scheduling into an FPGA. All of them are focused on FPGA area use optimization. Diessel and Elgindy [4], Bazargan [5], Handa [6] and Ahmadinia [7]. present interesting techniques to deal with task allocation in an FPGA that is being used for HW multitasking, but all of them neglect the fact that reconfiguration in present day FPGAs is a key aspect to make hardware multitasking effective. Another very interesting approach is presented in [9] which deals with the online problem of 2D task scheduling.

On the other hand, other authors focus on the reconfiguration aspect of the problem. Such is the case of Resano [10] who presents a technique that allows reducing reconfiguration overhead. This work is solving an offline problem, that is, the applications that are going to be executed are known as well as their execution graph. Thus the reconfiguration overhead may be hidden by conveniently scheduling the reconfiguration of the tasks in the graph. They are also working on several reconfigurable units and not a single FPGA and therefore they are not dealing neither with the area use part of the task scheduling problem nor with the only reconfigurable interface problem.

Götz [11] applies a server-based method and models the reconfiguration activities as aperiodic tasks. He applies the reconfiguration scheduling heuristics to the parts of the OS that are going to be executed in the reconfigurable hardware. Redaelli [12] has also dealt with the reconfiguration overhead problem and has presented techniques that increase task scheduling performance by including reconfiguration considerations into the task scheduling algorithm. However this author also solves the offline problem in 1D.

A very interesting work has been presented by Angermeier [13] in which he points at the importance of focusing both in the area use and reconfiguration port use in the same scheduling algorithm. They model the reconfiguration problem as a server-parallel machines problem in a very interesting way. They compare the reconfiguration port to the server and the slots defined in the FPGA to the parallel machines. This author uses the Erlangen Slot machine cited in [8] and includes some scheduling algorithms to the system. He presents three different heuristics, but two of them are only valid for two parallel machines. They are also working on the offline problem and they are using a 1D dynamic reconfiguration model. The FPGA they use is divided into m equal slots, and the reconfiguration time is equal for all the slots and is normalized to value 1.
To sum up, previous research in this area has been either focused on making a good FPGA area use by means of online scheduling heuristics, or focused on the reconfiguration scheduling for offline problems. This research work aims at solving reconfiguration scheduling issues for 2D online problems, which means that area availability in the FPGA has to be taken into account although it is not the main aspect under research.

## 3. SYSTEM MODEL

Our system is a hybrid general computing system where tasks may be executed in a software processor (microprocessor) or a hardware processor (FPGA). Tasks arrive at the hardware manager on demand, and it is never known when the next task will arrive for execution or how this task will be. Hardware multitasking is implemented by using 2D partial dynamic reconfiguration [27]. Most previous models consider the FPGA as a homogeneous CLB array. However, real FPGAs have BRAM blocks, multipliers and DSPs in a certain disposition and therefore this heterogeneity in the FPGA has to be taken into account for a realistic task allocation scheduling. This is especially important when the benchmarks being used are based on real hardware tasks. The restrictions in the possibilities for dynamic partial reconfiguration naturally points at the division of the FPGA into predefined slots, as suggested by [8], although in our case we are using this idea in a heterogeneous 2D FPGA.

From the synthetic benchmarks selected for our study, we have observed that there is a wide variety of FPGA resources required by different tasks, which adds an extra difficulty in task allocation in the FPGA. As area management was not our main objective, we decided to classify tasks according to the FPGA resources they need, so that the algorithms tested are focused on reconfiguration interface use.

Each task belongs to class $c_i$, the one that best fits task resources and thus fulfilling:

$$T_i \in c_i \text{ if resources } T_i \leq \text{ resources } c_i \tag{1}$$

Classes have been designed taking into account the number of CLBs, RAMs and DSPs available:

$$\sum_{i=0}^{N} \text{resources } c_i \leq \text{resources FPGA} \tag{2}$$

The number of classes is obtained, as pointed before, from a classification of the tasks expected for execution in the FPGA. If the sum of the number of resources needed by the classes is clearly lower than the resources available in the FPGA, then we may say that the differences in performance are related to the reconfiguration interface management. However, if the sum of the number of resources needed by the classes is larger than the resources available in the FPGA, this means that the FPGA is too small to execute the expected workload.

By combining the 2D slotted FPGA idea from [8] with a representative enough classification of synthetic tasks, we have been able to implement a simple FPGA area management in our algorithms, that is efficient enough to enhance the differences observed in hardware multitasking performance, due to the reconfiguration scheduling strategies under study.

FPGA area management is thus implemented by writing incoming tasks into the task queue associated to the task class, and may be scheduled for execution when a slot in the FPGA containing the resources needed by the task class becomes free as well as the reconfiguration port.

Tasks are written into queues which group them according to their class (Algorithm 1). Therefore, finding an empty slot that fulfils the task resource requirements for the task in the FPGA may be done with a fast and simple algorithm.

The N FPGA classes are named 0 to N-1, where class 0 represents tasks with the smallest resources' restrictions and N represents those with the highest FPGA resources needs (not just CLBs, but also DSPs and/or BRAMs). Subroutine *FitQj(Ti)* checks if Class j contains enough resources for $T_i$. Therefore Class$j$ is the smallest class where $T_i$ may be classified, and $T_i$ is written into waiting queue $Q_j$ associated to Class j (*Write(Ti,Qj)*).

```
int SelectQueue(Ti)
    j = 0;
    sel = false;
    while ((j < N-1) and (sel == false)) do
        if (FitQj(Ti) == true) then
Write(Ti,Qj)
            sel = true
            return = j;
              end if;
```

j++;
end while

Algorithm 1. Select Queue Algorithm

Where:

$$FitQ_j(Ti) = true \quad if$$

$$\begin{cases} \#CLBs\ of\ T_i \leq \#CLBS\ in\ class_j \\ \#BRAMs\ of\ T_i \leq \#BRAMs\ in\ class_j \\ \#DSPs\ of\ T_i \leq \#DSPs\ in\ class_j \end{cases} \quad (3)$$

In addition, before reconfiguring task Ti into the FPGA, the system has to check whether the time constraints for Ti (tmaxi) are met:

$$tmax_i \geq tstart_i + trec_i + tex_i \tag{4}$$

$tmax_i$: stands for the maximum time unit for the task to have finished execution
$trec_i$: stands for the reconfiguration time required by the task $T_i$
$tex_i$: stands for the execution time of the task $T_i$
$tstart_i$: stands for the time unit when $T_i$ will begin its reconfiguration

The values of $tmax_i$ and $tex_i$ are characteristic of the task. The value of $trec_i$ depends on the slot assigned to the class of the task, as reconfiguration time is proportional to the FPGA area reconfigured.

## 4. ORDERLY RECONFIGURATION STRATEGY

Orderly Strategy is the simplest idea to deal with the problem. This strategy uses a reconfiguration FIFO in order to guarantee that the reconfiguration scheduling is correct. Tasks are written into queues according to their sizes and their execution scheduling in the FPGA is done by taking into account both the availability of space in the FPGA as well as the availability of the reconfiguration port. The scheduling algorithm maintains task arrival order when assigning reconfiguration resources to tasks. This is a very simple algorithm with a low computational cost that we use as a baseline for the evaluation of the efficiency of the other algorithms presented.

When a task $T_i$ arrives for execution, an exact calculation for its starting time, $tstart_i$, is made, which depends on the availability of the reconfiguration interface and the availability of space in the FPGA:

tstart$_i$ = MAX[tfree$_{ICAP}$(T$_i$), tfree$_A$(T$_i$, Q$_j$), tarr$_i$]     (5)

In this equation, *tfreeICAP(Ti)* is the time unit at which the reconfiguration interface (Internal Configuration Access Port, ICAP) will be free to reconfigure another task and is calculated as follows:

tfree$_{ICAP}$(T$_i$)= tstart$_{i-1}$+trec$_{i-1}$     (6)

According to the Orderly Strategy, the task which arrived immediately before $T_i$, $T_{i-1}$, is the task scheduled to reconfigure before $T_i$. Therefore the reconfiguration interface will be free when $T_{i-1}$ finishes its reconfiguration.

In equation 5, *tfreeA( $Q_j$)* is the time unit at which there will be a free space for $T_i$, assigned to class j. This time is calculated by the area assignment strategy chosen for the implementation In our case, *tfreeA(Ti)* indicates the time unit when there is a free slot in the FPGA containing the resources needed by the Task Class j.

It is possible that both the reconfiguration interface and the FPGA be free at the arrival of Ti. In this case Ti is immediately reconfigured and executed: $tstart_i = tarr_i$, where $tarr_i$ stands for the arrival time of $T_i$.

The algorithm for the Orderly Strategy works as shown in algorithm 2. The function *Schedule(cj, Ti, tstarti)* schedules the reconfiguration of $T_i$ at time unit $tstart_i$. $T_i$ is deleted from queue $Q_j$, associated to class c at time tstarti, when its bitmap is loaded into the FPGA. The function *Reject(Ti)* deletes $T_i$ from the queue and reports that it is not possible to execute it in the FPGA with the present constraints.

```
tstart₁ = 0;
tfreeA(T1) = 0;
tfreeICAP(T1) = 0;
while (tasks in queues)
        j = SelectQueue(Ti);
        tstarti = MAX[tfreeICAP(Ti), tfreeA(Qj), tarrivali];
        if (tmaxi   tstarti+treci+texi)
      Schedule(cj, Ti, tstarti);
   else
      Reject(Ti);
        end if;
        i ++;
end while;
```

Algorithm 2. Orderly Strategy, reconfiguration scheduling algorithm

Therefore, the Orderly Strategy may be applied when a task Ti arrives at the FPGA. Its start time can be calculated accurately, so whether the task may be executed within its time constraints or not is immediately known. Tasks are reconfigured at the tstarti time scheduled into the reconfiguration FIFO, which follow the same arrival order of tasks. By using this strategy there are intervals of time when both the reconfiguration interface and there is enough free space in the FPGA, but not for the next task scheduled for reconfiguration. These intervals could be used to reconfigure tasks assigned to the free slots and a more efficient use of the reconfiguration interface and FPGA area would be achieved.

## 5. RECONFIGURATION PORT INTENSIVE USE STRATEGY

As we have seen, there is only one reconfiguration interface for the whole FPGA, which may reconfigure any of the first tasks in each queue when the ICAP becomes free. The overall scheduling problem may be viewed as a multiple stations-one server problem, where the server is the reconfiguration interface and the stations are the tasks' queues associated with free FPGA slots of a given class.

When a new task arrives, the scheduling algorithm assigns a queue for the task according to its characteristics In order to make an intensive use of the ICAP, the reconfiguration time for this task will be decided later, so that when the ICAP becomes free any task from the queues' heads may be chosen for reconfiguration, as long as there is enough free space in the FPGA.

Function *FreeA(Qc)=(tfreeA, Tj)* returns $T_j$, which is the task waiting for execution in queue $Q_c$, associated to class c, and time unit tfreeA is the time when a FPGA slot with the characteristics of class c will be free.

Function *StrategyRPIU(FreeQ(0), FreeQ(1), …)* is executed each time there are free slots in the FPGA as well as the reconfiguration interface is available. This function analyzes the values returned by *FreeA()* for each class and chooses that with the lowest *tfreeA* and the task associated that will be scheduled next for reconfiguration and execution only if equations 4 is fulfilled. Otherwise, that task is rejected and deleted from the queue and function *StrategyRPIU()* is executed again in search for a new candidate (shown in algorithm 3).

$tstart_1=0$;
$tfree_A(T_1)=0$;
$tfree_{ICAP}(T_1)=0$;
if (reconfiguration interface = free)
$[(Q_i, T_i), (Q_k, T_v), …]=$ **StrategyRPIU**(FreeA(0), FreeA(1), …);
$(Q_j, T_i) =$ **ICAParbiter**$((Q_j, T_i), (Q_k, T_v), …)$;
$tstart_i =$ MAX$[t, tfree_{ICAP}(T_k), tfree_A(Q_j)]$;
    if $(tmax_i \quad tstart_i+trec_i+tex_i)$
       **Schedule**$(Q_j, T_i, tstart_i)$;
    else
      **Reject**$(T_i)$;
    end if;
end if;

Algorithm 3. Reconfiguration Port Intensive Use Strategy, reconfiguration scheduling algorithm

It may happen that more than one class finds available slots in the FPGA at the same time unit (or that the only free slot is suitable for several classes); in such cases, the function *StrategyRPIU* returns one task for each class for which there is available space in the FPGA. Then the *ICAParbiter()* chooses which task is first to be reconfigured.

The starting time for $T_i$ is calculated as shown the algorithm, where: t is the current time unit and fulfills t tarr$_i$, *tfreeICAP($T_k$)= tstart$_k$ + trec$_k$* where $T_k$ is the task scheduled to reconfigure before $T_i$ and *tfreeA($Q_j$)=tstart$_m$+trec$_m$+tex$_m$* is the time when a slot with the resources for class $Q_i$ will be free ($T_m$ is the task scheduled to execute in this slot before $T_i$).

This strategy allows making an intensive use of the reconfiguration port and therefore yields better performance results than the basic orderly strategy.

We have developed two distinct policies that implement the RPIU strategy in combination with two priority criteria that select the task to be reconfigured and executed first in cases where there is a reconfiguration interface use conflict, that is, two different version of the ICAParbiter() policy.

## 5.1 Most Loaded Queue Policy

This algorithm implements the RPIU Strategy with a version of the ICAP arbiter function that gives priority to tasks in the most loaded queue (algorithm 4).

This version of the *ICAParbiter()* first checks if the task in the queue head fits the free slot; it then checks if this queue is more loaded than the other queues containing tasks that also fit the free slot. When it has reviewed all queues, the function returns the task written at the head of the most loaded queue that fits the free slot. If there are no tasks that fit the free slot, the function returns T = null.

Function *Fit(free_slot,$T_i$)* checks that the free slot has enough resources for the execution of Ti, as shown in equation 2. Function *load($Q_j$)* returns the number of tasks in $Q_j$. The variable old_sel is used to avoid that one of the classes may become too greedy and block access to FPGA resources to the other classes. The use of this variable allows this policy to occasionally reconfigure tasks from less loaded queues in-between the reconfiguration of tasks from the most loaded queue.

```
function  ICAParbiter(Qj, Qk, …)
sel = null;
for j=0 to N-1
        T = head(Qj);
        if Fit(free_slot,T) = true
                if load(sel) < load(Qj)
                        if old_sel    Qj
                                sel = Qj;
                        end if;
                end if;
        end if;
end for;
if sel=null
        old_sel=null;                // the only task that fits free slots is from old_sel
        (sel, T)= ICAParbiter(Qj, Qk, …)
end if;
old_sel=sel;
return sel, T;
```

Algorithm 4. ICAP Arbiter Function for Most Loaded Queue Policy

## 5.2 Deadline Policy

This ICAP arbiter function (Algorithm 5) gives priority to the task with the lowest execution margin. It is focused on reconfiguring those tasks which are closer to be rejected.

When tmargin is the same for more than one task, the Deadline Policy algorithm chooses the task that belongs to the class which at this time makes the best use of the FPGA free resources.

```
function ICAParbiter(Qj, Qk, …)
sel = null;
T = null;
tmargin=10¹⁵;
for j=0 to N-1
        Ti = head(Qj);
        if Fit(free_slot,Ti) = true
                if tmaxi – (t + treci +texi) < tmargin;
                        sel = Qj;
                        T = Ti;
                        tmargin = tmaxi – (t + treci + texi);
                                else if tmaxi – (t + treci +texi) = tmargin;
                                        if compare_fit(T, Ti, free_slot) = true
                                                sel = Qj;
```

tmargin = tmax$_i$ – (t + trec$_i$ +tex$_i$);
       end if;
            end if;
    end if;
 end for;
**return** sel, T;

Algorithm 5. ICAP Arbiter Function for Deadline Policy

This version of the *ICAParbiter()* first checks there is more than one task that fit the slot with the minimum tmargin, the task that makes the best use of the resources in the slot is selected. This is implemented by using function compare_*fit(T, T$_i$, free_slot)*, which returns a true value if Ti makes better use of the free slot than T. When *ICAParbiter()* has reviewed all queues, it returns the task (among those in the head of queues) with the lowest deadline that best fits the free slot, as well as its associated queue. If there are no tasks that fit the free slot, the function returns T = null.

The criterion used to define a better use of the resources in the slot makes use of the following equation:

pairs(x) = #CLBs·(#LUT-Flipflop_pair/CLB) +
          #BRAMs·(#LUT-Flipflop_pair/CLB) +          (7)
          #DSPs·(#LUT-Flipflop_pair /CLB)

where x may either be a task or a slot. We are using this definition because all FPGA vendors provide the correspondence between pair LUT-flip-flop and the different elements in their FPGAs as standard FPGA size unit. Therefore, if compare_fit(T, Ti, free_slot) is true it means that the value of pairs(Ti) is closer to pairs(free_slot) than pairs(T).

## 6. EXPERIMENTAL RESULTS

We have developed several experiments in order to compare the performance of the different algorithms presented. As was previously explained, the objective of this paper is to discuss the impact of the reconfiguration strategy on the performance of a hardware multitasking scheduling algorithm.

We have used information from real applications' hardware synthesis from Xilinx in order to develop several synthetic benchmarks that are described in detail in subsection 6.3.
We have also generated artificial sets of tasks with a balanced profile in relation to the classes defined from the real application data.

### 6.1 Experiments' design

For each set of tasks we have done an offline study and have calculated a theoretical average rejection ratio of the tasks in the set, ARR. As we have explained, each task is restricted by the value of its tmax, and the chances to reject the task are related to the ratio between the offline estimation of the task execution end time and its tmax:

$$Reject\ ratio_i = \frac{tsart_i + trec_i + tex_i}{tmax_i} \qquad (8)$$

Therefore the higher ARR of a task set is, the lower we expect the number of executed tasks to be. Our study includes benchmarks with values of ARR over 1, because we are assuming a regular intensive use of the FPGA. And it is also possible that at some time intervals our system may need to execute even more computational work on the FPGA. Therefore, it is very important to understand how the different strategies behave in heavy workload situations.

A set of tasks consists of a file with 52 different tasks arranged according to their arrival times. This simulates the online arrival of tasks at the hardware manager. The hardware manager reads one task from the file and does not have any information about the characteristics of the rest of the tasks in the file (arrival of next task, number of remaining tasks...) as happens in an online system.

We are using the ARR factor to characterize our experiments. We have obtained sets of tasks with a low ARR by either leaving long time intervals between the arrival of tasks or by assigning them a very high tmax value (or a combination of both). As was previously discussed, for the sets with a low ARR, the impact on the value of tstart due to the reconfiguration interface is very low because most of the new arriving tasks will find the reconfiguration interface free. In addition, if the tasks' deadlines are high, they can afford to wait for a long time until they may be reconfigured. Then, the sets with a low ARR should be less sensitive to the reconfiguration strategy used and we would expect to execute 100% of tasks for any of the reconfiguration strategies used. If either the arrival times of tasks are very close to each other or the tmax value of the tasks in the set is small (only a little margin for reconfiguration and execution) or a combination of both, then the set of tasks has a high ARR and it is expected that many tasks will be rejected. In these situations, a good management of the reconfiguration interface becomes crucial and therefore these sets of tasks are more sensitive to the reconfiguration strategy used. This is the reason why we have focused our experiments on sets of tasks with an ARR value around 1.

## 6.2 Description of the target architecture

We have used a simulation platform with a Virtex 5 XC5VSX50T FPGA according to the technical details provided by Xilinx in [26]. This FPGA consists of 4080 CLBs, 132 BRAMs and 288 DSPs.

Four different tasks' classes have been defined according to the analysis of the data collected from Xilinx cores of real applications, as shown in Table 1. In a hardware multitasking system it is not expected that tasks executing will need the whole FPGA resources for execution, nor that most of them will need less than 10% of it. We believe that these four classes of tasks chosen for our benchmarks is representative enough of the hardware multitasking environment and serves the purpose of our study which is the impact of the reconfiguration scheduling on total system performance. The reconfiguration times associated to each task class have been normalized. Thus, the basic time unit used for the experiments is the reconfiguration time for the smallest task assigned present in C0.

Table 1. Task class profiling

| Task class | max. #CLBs | max. #BRAMs | max. #DSPs |
|:---:|:---:|:---:|:---:|
| $C_0$ | 204 | 4 | 16 |
| $C_1$ | 480 | 20 | 32 |
| $C_2$ | 1000 | 18 | 80 |
| $C_3$ | 2400 | 90 | 160 |

## 6.3 Descriptions of the benchmarks

We have created three groups of Balanced Artificial benchmarks in which the fraction of the workload according to tasks' classes (sum of reconfiguration and execution times of the tasks for each class) are equal. Thus the workload is perfectly balanced for the resources of the FPGA. We have created 10 sets of tasks with different ARR values for each group (see Table 2). The value of the arrival time of the last task in the set is also shown to highlight whether the differences in the ARR are due to an increase in task frequency rate.

Table 2. Perfect, Semi-perfect and Global Balanced Benchmark Profiling

| PB | ARR | tarr$_{52}$ | SB | ARR | tarr$_{52}$ | GB | ARR | tarr$_{52}$ |
|-----|------|------|-----|------|------|-----|------|------|
| PB0 | 0.85 | 260 | SB0 | 0.85 | 260 | GB0 | 0.85 | 260 |
| PB1 | 0.88 | 258 | SB1 | 0.88 | 258 | GB1 | 0.88 | 258 |
| PB2 | 0.91 | 258 | SB2 | 0.91 | 258 | GB2 | 0.91 | 248 |
| PB3 | 0.94 | 258 | SB3 | 0.94 | 210 | GB3 | 0.94 | 200 |
| PB4 | 0.98 | 206 | SB4 | 0.98 | 210 | GB4 | 0.98 | 200 |
| PB5 | 1.00 | 206 | SB5 | 1.00 | 188 | GB5 | 1.00 | 188 |
| PB6 | 1.03 | 206 | SB6 | 1.03 | 188 | GB6 | 1.03 | 188 |
| PB7 | 1.06 | 154 | SB7 | 1.06 | 188 | GB7 | 1.06 | 198 |
| PB8 | 1.10 | 154 | SB8 | 1.10 | 136 | GB8 | 1.10 | 136 |
| PB9 | 1.15 | 154 | SB9 | 1.15 | 136 | GB9 | 1.15 | 126 |

- Perfect Balance (PB): the arrival order of tasks is uniform in relation to their classes, that is, a task from C0 arrives first, then a task from C1, then another from C2, etc.

- Semi-perfect Balance (SB): the arrival order of 25% of tasks in the PB task set has been modified.

- Global Balance (GB): although the total workload is perfectly balanced in terms of tasks' classes, the arrival order of tasks is random. However, we have checked that no more than three tasks of the same class arrive consecutively.

We have created three groups of synthetic benchmarks, S1, S2 and S3, using Xilinx cores for different real applications. Table 3 describes the three groups of synthetic benchmarks created from Xilinx cores for different applications available at: xilinx.com/support/documentation/ip_documentation/. For each group of synthetic benchmarks we have designed three different sets of tasks. Their profiling is shown in table V. As these sets of tasks are synthetic, we have not been able to create perfectly balanced profiles with them. Therefore we also include information about the workload profiling, where the fraction of tasks of each class in the total workload has been represented.

Table 3. Composition of Synthetic Benchmarks

| Application | S1 | S2 | S3 |
|---|---|---|---|
| Defective pixel correction [14] | 7 | - | 7 |
| Gamma correction I [15] | 6 | - | 8 |
| Gamma correction III [15] | - | 5 | 6 |
| Image processing pipeline [16] | - | - | 6 |
| Colour correction matrix [17] | - | 6 | 7 |
| Reed solomon encoder [18] | 8 | 6 | - |
| Reed solomon decoder [19] | 8 | 6 | - |
| Convolutional encoder [20] | 6 | 14 | - |
| Viterbi Decoder [21] | - | 4 | - |
| UMTS Decoder I [22] | - | 4 | - |
| UMTS Decoder III [22] | - | - | 6 |
| CORDIC cosin [23] | - | - | 6 |
| DFT [24] | 1 | - | - |
| FFT [25] | 16 | 7 | 6 |
| **Total** | **52** | **52** | **52** |

Table 4. Profiles of Synthetic Benchmarks

| | **ARR** | **tarr$_{52}$** | **C$_0$** | **C$_1$** | **C$_2$** | **C$_3$** |
|---|---|---|---|---|---|---|
| S1A | 0.75 | 242 | 0.24 | 0.25 | 0.24 | 0.27 |
| S2A | 0.75 | 205 | 0.21 | 0.28 | 0.28 | 0.26 |
| S3A | 0.76 | 303 | 0.13 | 0.31 | 0.30 | 0.26 |
| S1B | 0.81 | 187 | 0.24 | 0.25 | 0.24 | 0.27 |
| S2B | 0.80 | 201 | 0.21 | 0.28 | 0.28 | 0.26 |
| S3B | 0.80 | 303 | 0.13 | 0.31 | 0.30 | 0.26 |
| S1C | 0.88 | 184 | 0.24 | 0.25 | 0.24 | 0.27 |
| S2C | 0.87 | 154 | 0.21 | 0.28 | 0.28 | 0.26 |
| S3C | 0.88 | 252 | 0.13 | 0.31 | 0.30 | 0.26 |

## 6.3 Simulation Results

In this subsection we present four tables, one for each group of benchmarks. The first column shows the name of set of task, the second column shows the ARR obtained for these sets of tasks and the other columns, divided into two sub-columns each, show the performance results of the two strategies and their associated policies.

We are using two parameters to describe the performance of a scheduling strategy: n, the total number of executed tasks by the strategy, and improvement, which represents the performance ratio of the RPIU strategy, and its two policies, in comparison with the Orderly Strategy (percentage of tasks executed by any strategy over those executed by Orderly Strategy).

Table 5 shows that for Perfect Balanced benchmarks all algorithms execute all tasks for ARR values up to 0.94. This is due to the uniformity of these sets of tasks, as the balance of their workload is ideal. For higher values of ARR, none of the strategies is able to execute all tasks, as expected, and we observe that the RPIU with Deadline Policy always yields better performance than any of the other two.

Table 5. Results for Perfect Balanced Benchmarks

| | ARR | Orderly | Deadline | | Queue Load | |
|---|---|---|---|---|---|---|
| | | n | n | Improvement | n | Improvement |
| PB0 | 0.85 | 52 | 52 | 0 % | 52 | 0 % |
| PB1 | 0.88 | 52 | 52 | 0 % | 52 | 0 % |
| PB2 | 0.91 | 52 | 52 | 0 % | 52 | 0 % |
| PB3 | 0.94 | 52 | 52 | 0 % | 52 | 0 % |
| PB4 | 0.98 | 50 | 50 | 0 % | 50 | 0 % |
| PB5 | 1 | 49 | 50 | 1.9 % | 50 | 1.9 % |
| PB6 | 1.03 | 47 | 47 | 0 % | 48 | 1.9 % |
| PB7 | 1.06 | 32 | 45 | 25 % | 46 | 26.9 % |
| PB8 | 1.10 | 36 | 44 | 15.4 % | 45 | 17.3 % |
| PB9 | 1.15 | 30 | 40 | 19.2 % | 41 | 21.1 % |

Table 6. Results for Semi-Perfect Balanced Benchmarks

| | ARR | Orderly | Deadline | | Queue Load | |
|---|---|---|---|---|---|---|
| | | n | n | Improvement | n | Improvement |
| SB0 | 0.85 | 52 | 52 | 0 % | 52 | 0 % |
| SB1 | 0.88 | 51 | 52 | 1.9 % | 52 | 1.9 % |
| SB2 | 0.91 | 50 | 52 | 2.8 % | 52 | 2.8 % |
| SB3 | 0.94 | 43 | 52 | 17.1 % | 52 | 17.1 % |
| SB4 | 0.98 | 43 | 48 | 9.5 % | 48 | 9.5 % |
| SB5 | 1 | 38 | 49 | 20.9 % | 48 | 20.9 % |
| SB6 | 1.03 | 38 | 48 | 19 % | 47 | 17.1 % |
| SB7 | 1.06 | 38 | 45 | 13.4 % | 44 | 11.5 % |
| SB8 | 1.10 | 29 | 40 | 21.1 % | 40 | 21.1 % |
| SB9 | 1.15 | 29 | 36 | 13.4 % | 36 | 13.4 % |

From Table 6 we observe that although the sets of tasks are also balanced, the differences in the performance of the reconfiguration algorithms become noticeable for lower values of ARR than in the previous case. This group of experiments shows again that the RPIU Strategy with Deadline Policy is better than the others. Although for the experiment SB5, the number of executed tasks with the Queue Load Policy is slightly higher than that of the Deadline. Both policies achieve better results than the Orderly Strategy from an ARR value of 0.91.

Table 7 follows the trend seen in Table 6. For this group of experiments, a high value of ARR has more impact on the number of executed tasks than we have seen in the previous experiments. These experiments have been designed with random arrival of tasks and therefore there are lapses of time when the workload assigned to the FPGA is largely over the FPGA capacity. The more efficient the reconfiguration strategy is, the lower the impact of these task peaks on the global performance. We may see that both Queue Load and Deadline policies yield good performance for limit ARR values (all 52 tasks are executed even with ARR=0.94) while the Orderly Strategy starts rejecting tasks for ARR values over 0.85.

We would like to remark that Table 8 presents three different synthetic benchmarks with three different values for ARR each, which have been grouped in contiguous lines for clarity. These benchmarks are formed by tasks with a wide variety in resource requirements as well as in execution times: a Discrete Fourier Transform takes about a hundred times more time to execute than a Color Correction Matrix and uses about 30 times more FPGA resources. In spite of this, both the Deadline Strategy and the Queue Load Policy are able to schedule all tasks for execution for an ARR value of 0.85, and fail to schedule 100% of the tasks for higher values of ARR. However, in all cases the performance of the RPIU Strategy is well over the performance of the Orderly.

Table 7. Results for Global Balanced Benchmarks

|  | ARR | Orderly | Deadline | | Queue Load | |
|---|---|---|---|---|---|---|
|  |  | n | n | Improvement | n | Improvement |
| GB0 | 0.85 | 47 | 52 | 5.7 % | 52 | 5.7 % |
| GB1 | 0.88 | 44 | 51 | 13.4 % | 51 | 13.4 % |
| GB2 | 0.91 | 43 | 50 | 13.4 % | 50 | 13.4 % |
| GB3 | 0.94 | 46 | 50 | 7.7 % | 50 | 7.7 % |
| GB4 | 0.98 | 44 | 50 | 11.5 % | 50 | 11.5 % |
| GB5 | 1 | 44 | 48 | 7.7 % | 49 | 7.7 % |
| GB6 | 1.03 | 41 | 43 | 2.8 % | 42 | 1.9 % |
| GB7 | 1.06 | 39 | 40 | 1.9 % | 40 | 1.9 % |
| GB8 | 1.10 | 36 | 40 | 7.7 % | 40 | 7.7 % |
| GB9 | 1.15 | 30 | 40 | 19 % | 41 | 17.3 % |

Table 8. Result for Synthetic Benchmarks

|  | ARR | Orderly | Deadline | | Queue Load | |
|---|---|---|---|---|---|---|
|  |  | n | n | Improvement | n | Improvement |
| S1A | 0,85 | 42 | 52 | 19 % | 52 | 19% |
| S2A | 0,85 | 49 | 52 | 5.7 % | 52 | 5.7 % |
| S3A | 0,86 | 52 | 52 | 0 % | 52 | 0% |
| S1B | 0,91 | 41 | 48 | 13.4 % | 47 | 11.5 % |
| S2B | 0,91 | 36 | 41 | 9.6 % | 40 | 7.7 % |
| S3B | 0,91 | 51 | 51 | 0% | 51 | 0% |
| S1C | 0,98 | 39 | 47 | 15.3 % | 46 | 13.4  % |
| S2C | 0,97 | 27 | 34 | 13.4 % | 36 | 17.1 % |
| S3C | 0,98 | 43 | 43 | 0% | 42 | -1.9% |

This group of experiments is most interesting, not only because it simulates the behavior of real applications, but also because of the difficulties that the dispersion in both the execution times and the tasks' sizes present, and highlights the importance of a good reconfiguration strategy in hardware multitasking scheduling.

We have calculated the average number of tasks executed for all benchmarks (artificial and synthetic) grouped by their ARR value, and have represented them in a graph shown in Figure 2. Being conscious that although we have made a wide range of experiments, we can never cover the infinite possibilities that may be found in a real computing system, we have calculated a linear interpolation of the results in order to find out whether there is a clear difference in the performance tendency of the strategies under study.

The analysis of this graph clearly points at RPIU Strategy with Deadline Policy as a better reconfiguration management than the Orderly Strategy and the RPIU with Queue Load policy. In systems where an intensive use of the FPGA is expected (ARR > 0,8) the reconfiguration strategy used is crucial for a good performance and the Deadline Poicy is the most suitable of the three strategies analysed.

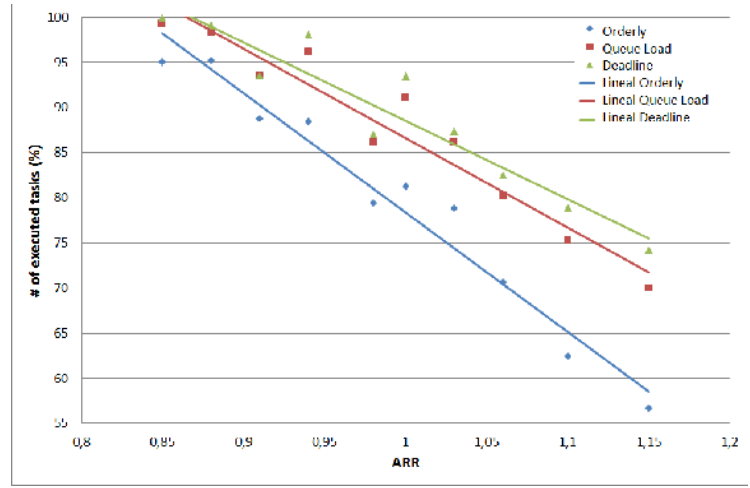

Figure 2. Lineal approximation of the number of tasks executed by the strategies tested

# 6. CONCLUSION

In this paper we have presented different reconfiguration strategies which have been combined with a simple heuristic used to find an empty slot in the FPGA.

The FIFO strategy schedules the start time of the tasks as soon as they arrive. Then the FPGA manager may report task rejection at a very early stage.

The RPIU strategy delays this scheduling to the time when a task may be executed in the FPGA, this is, when there is a free slot in the FPGA and the reconfiguration interface is free. Delaying the scheduling decision allows taking into account more parameters that are available in the system at task launch time. Therefore, RPIU uses a wider scope for the scheduling and yields better results. The performance difference between the RPIU policies is not as remarkable as that of RPIU and FIFO. The difference lies in the priority criteria used to resolve reconfiguration conflicts (there are enough FPGA resources available for several tasks at the same time).

The aim of the Queue Load Policy is to keep the workload balance among the queues. In case of conflict this policy gives priority to the task in the most loaded queue. The Deadline Policy calculates the execution margin of tasks in conflict and gives priority to the one with the lowest remaining time.

The experimental results show that when the workload assigned is close to the maximum theoretical FPGA capacity, the RPIU strategy achieves better results. Among the two RPIU policies, the Deadline is the one that achieves more efficient results in most experiments. The explanation for this is that if in times of conflict there is a queue that is significantly more loaded than the rest, it is very likely that the tasks in this queue may have been waiting for a long time. Therefore, the Deadline Policy will launch them for execution, similar to what Queue Load would do. However, if the tasks in the most loaded queue may wait for a longer time than the others, it is reasonable to give priority to tasks that are closer to rejection. In addition, as displayed in Fig. 2, the RPIU Strategy with Deadline Policy has a tendency to make the best use of the FPGA resources.